\documentclass[twocolumn,pra]{revtex4}

\usepackage{graphicx}
\usepackage{bm}
\usepackage{amsmath}

\begin{document}

\title{Decoherence of cold atomic gases in magnetic micro-traps}

\author{C. Schroll}\author{W. Belzig}\author{C. Bruder}
\affiliation{Department of Physics and Astronomy, University of Basel, 
Klingelbergstrasse 82, CH-4056 Basel, Switzerland}

\date{\today}

\begin{abstract}
We derive a model to describe decoherence of atomic clouds in
atom-chip traps taking the excited states of the trapping potential
into account. We use this model to investigate decoherence for a
single trapping well and for a pair of trapping wells that form the
two arms of an atom interferometer. Including the discrete spectrum of the
trapping potential gives rise to a decoherence mechanism with a decoherence
rate $\Gamma$ that scales like $\Gamma \sim 1/r_0^4$ with the distance $r_0$
from the trap minimum to the wire.
\end{abstract}


% 03.75.Dg Atom and neutron interferometry
% 03.75.-b matter waves
% 03.75.Gg Entanglement and decoherence in Bose-Einstein condensates
% 03.65.Yz Decoherence; open systems; quantum statistical methods 

\maketitle

%%%%%%%%%%%%%%%%%%%%%%%%%%%%%%%%%%%%%%%%%%%%%%%%%%%%%%%%%%%%%%%%%%%%%%
\section{\label{sec:level1}Introduction}
Cold atomic gases form an ideal system to test fundamental quantum
mechanical predictions. Progress in laser cooling made it possible to achieve
previously inaccessible low temperatures in the nK-range (see, e.g.,
the Nobel lectures 1998 \cite{nobel98}). One of the most exciting
consequences of this development has been the creation of 
Bose-Einstein condensates (BECs) \cite{Anderson01,Davis01}. These have
been manipulated by means of laser traps in various manners, e.g. vortices
have been created or collisions of two BECs have been studied
\cite{Madison01,Abo-Shaeer01,Andrews01}. An interesting link to solid-state
phenomena has been established by creating optical lattices, in which a
Mott transition has been theoretically predicted \cite{Jaksch01} and observed
\cite{Greiner01}.

Recently, proposals to trap cold atomic gases using microfabricated structures
\cite{weinstein} have been realized experimentally
\cite{prentiss,Denschlag01,Reichel01} on silicon substrates, so-called atom
chips. These systems combine the quantum mechanical testing ground of quantum
gases with the great versatility in trapping geometries offered by the
micro-fabrication process. Micro-fabricated traps made it possible to split
clouds of cold atomic gases in a beam splitter geometry \cite{Cassettari01},
to transport wavepackets along a conveyer belt structure \cite{Haensel01} and
to accumulate atomic clouds in a storage ring \cite{Sauer01}. Moreover, a BEC
has been successfully transferred into a micro-trap and transported along a
waveguide created by a current carrying micro-structure fabricated onto a chip
\cite{Haensel02,Ott01,Schneider01,Leanhardt01}. Finally several suggestions to
integrate an atom interferometer for cold gases onto an atom chip have
been put forward \cite{Hinds01,Andersson01,Haensel03}.

The high magnetic-field gradients in atom-chip traps provide a
strongly confined motion of the atomic quantum gases along the
micro-structured wires. The atomic cloud will be situated in close
vicinity of the chip surface. As a consequence, there will be
interactions between the substrate and the trapped atomic cloud, and
the cold gas can no longer be considered to be an isolated system.
Recent experiments reported a fragmentation of cold atomic clouds or
BECs in a wire waveguide
\cite{Leanhardt01,Leanhardt02,Fortagh01,Kraft01} on reducing the
distance between the wavepackets and the chip surface, showing that
atomic gases in wire traps are very sensitive to its
environment. Experimental \cite{Jones01} and theoretical works
\cite{Henkel03} showed that there are losses of trapped atoms due to
spin flips induced by the strong thermal gradient between the atom
chip, held at room temperature, and the cold atomic cloud. Moreover,
Refs. \cite{Henkel01,Henkel02} studied the influence of magnetic 
near-fields and current noise on an atomic wavepacket. Consequences
for the spatial decoherence of the atomic wavepacket were discussed under
the assumption that the transversal states of the one dimensional-waveguide
 are frozen out.

For wires of small width and height, as used for atom-chip traps, 
the current fluctuations will be directed along the wire. Consequently,
the magnetic field fluctuations generated by these current fluctuations, 
are perpendicular to the wire direction,
 inducing transitions between different transverse trapping states.

In this paper we will discuss the influence of current fluctuations
in micro-structured conductors used in atom-chip traps. We derive an
equation for the atomic density matrix, that describes the decoherence and
equilibration effects in atomic clouds taking transitions between different
transversal trap states into account. We study decoherence of an atomic state
in a single waveguide as well as in a system of two parallel waveguides.
Decoherence effects in a pair of one-dimensional (1d) waveguides are of
particularly high interest as this setup forms the basic building block for an
on-chip atom interferometer \cite{Hinds01,Andersson01,Haensel03}.

Our main results can be summarized as follows. Considering current
fluctuations along the wire, we show that spatial decoherence along
the guiding axis of the 1d-waveguide occurs only by processes
including transitions between different transversal states. The
decoherence rate obtained scales with the wire-to-trap distance $r_0$
as $\Gamma\sim 1/r_0^4$. Applying our model to a double waveguide
shows that correlations among the magnetic-field fluctuations in the
left and right arm of the double waveguide are of minor importance.
The change of the decoherence rate under the variation of the distance
between the wires is dominated by the geometric rearrangement of the
trap minima.

To arrive at these conclusions we proceed as follows. Section \ref{sec1} will
give a derivation of the kinetic equation describing the time evolution of an
atomic wavepacket in an array of $N$ parallel 1d waveguides subject to
fluctuations in the trapping potential. We will then use this equation of
motion in Section \ref{sec2} to study the specific cases of a single 1d
waveguide and a pair of 1d waveguides. Spatial decoherence and equilibration
effects in the single and double waveguide will be discussed. Finally we
summarise and give our conclusions in Section~\ref{sec3}.

%%%%%%%%%%%%%%%%%%%%%%%%%%%%%%%%%%%%%%%%%%%%%%%%%%%%%%%%%%%%%%%%%%%%%%
\section{\label{sec1}Equation for the density matrix}

The trapping potential of a micro-structured chip is produced by the
superposition of a homogeneous magnetic field and the magnetic field induced
by the current in the conductors of the micro-structure. Atoms in a low-field
seeking hyperfine state $|S\rangle$ will be trapped in the magnetic field
minimum \cite{Leggett01}. The interaction of the atom with the magnetic field
 is
\begin{equation}
V({\bf x})=-\langle S|\bm{\mu}|S
\rangle {\bf B}({\bf x}) \; ,
\label{trappot}
\end{equation}
assuming that the Larmor precession of the trapped atom is much faster than
the trap frequency and, hence, the magnetic moment of the atom
$\bm{\mu}$ can be replaced by its mean value 
$\langle S|\bm{\mu}|S\rangle$.

\begin{figure}
\includegraphics[width=6cm]{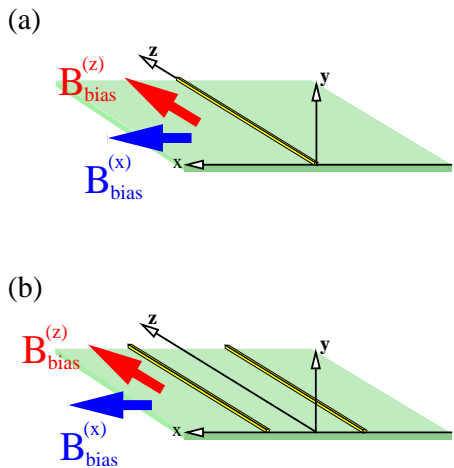}
\caption{Setup of the atom chip showing directions of the wires
 and the magnetic bias fields needed to form the trapping potential.
(a) Single-wire trap. (b) Double-wire trap.\label{figure0}}
\end{figure}

\begin{figure}
 \includegraphics[width=8.5cm]{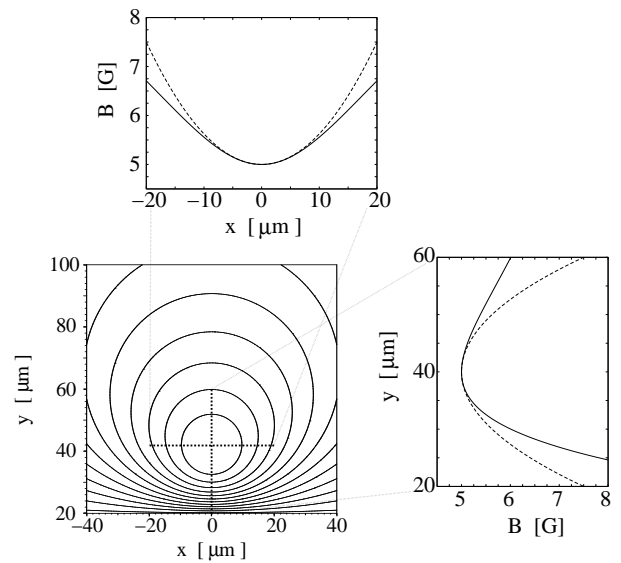}
 \caption{Contour-plot of the magnetic field of the single-wire trap
 for 
 $B^{(x)}_{\mathrm{bias}}= 10{\mathrm G}$, $B^{(z)}_{\mathrm{bias}}=
 5{\mathrm G}$, and $I=0.1{\mathrm A}$. The upper (right) plot shows a
 horizontal (vertical) cut through the potential minimum. The dashed lines
 in the upper and the right plot show the harmonic approximation to the
 trapping field. \label{figure2}}
\end{figure}

Many different wire-field configurations will lead to an atom trap
\cite{Reichel02,Folman01}. We will concentrate on systems in which the
trapping field is produced by an array of parallel wires, see
Fig.~\ref{figure0} for a single or double-wire trap. A homogeneous
bias field is applied parallel to the surface on which the wires are
mounted. An example for the resulting field distribution for the single-wire
trap is shown in Fig.~\ref{figure2}. 

%%%%%%%%%%%%%%%%%%%%%%%%%%%%%%%%%%%%%%%%%%%%%%%%%%%%%%%%%%%%%%%%%%%%%%
\subsection{\label{sec1a} Stochastic equation for the density matrix}

The quantum mechanical evolution of a cold atomic cloud is described by a
density matrix. The time evolution of the density matrix
$\rho({\bf x},{\bf x}^\prime,t)$ is given by the von Neumann equation
\begin{equation}
i\hbar\frac{\partial}{\partial t}\rho = [H,\rho] \;,
\end{equation}
where $H$ is the Hamiltonian of an atom in the trapping potential:
\begin{equation}
H = \frac{{\bf p}^2}{2m} + V_t({\bf r}_\perp)
 +\delta V({\bf x},t)\label{ham} \; .
\end{equation} 
Here, $V_t({\bf r}_\perp)$ is the confining potential which we will 
assume to be constant along the trap, and $\bf r_{\perp}$ denotes
the coordinate perpendicular to the direction of the current-carrying
wire or the wave guide. The last term $\delta V({\bf x},t)$ is a
random fluctuation term induced by the current noise.

We will formulate the problem for a system of $N$ parallel quasi-1d
magnetic traps generated by a set of $M$ parallel wires on the chip
(in general, $N\ne M$ since some of the magnetic-field minima may
merge). The number of parallel trapping wells is included in the
structure of the confining potential $V_t({\bf r}_{\perp})$. In all
further calculations we will assume that the surface of the atom chip
is in the $\hat{x}$-$\hat{z}$ plane and that the atoms are trapped in
the half space $y>0$ above the chip. The wires on the atom chip needed
to form the trapping field, are assumed to be aligned in
$\hat{z}$-direction.

In position representation for the density matrix the von
Neumann equation reads
\begin{eqnarray}
 i\hbar\frac{\partial}{\partial t}&\rho({\bf x},{\bf x}^\prime,t) = 
 \Big[-\frac{\hbar^2}{2m}\left(\frac{d^2}{d{\bf x}^2} -
 \frac{d^2}{d{\bf x}^{\prime^2}}\right)
 +V_t({\bf r}_{\perp}) \label{3vN} \\\nonumber
 &-V_t({\bf r}_{\perp}^\prime) +\delta V({\bf x},t) 
 - \delta V({\bf x}^\prime,t)\Big] 
 \rho({\bf x},{\bf x}^\prime,t)\; .
\end{eqnarray}
To derive a quasi-1d expression for Eq.~(\ref{3vN}) we
expand the density matrix in eigenmodes of the transverse potential
$\chi_n({\bf r}_{\perp})$:
\begin{equation}
 \rho({\bf x},{\bf x}^\prime,t) = \sum_{n,m}
 \chi_n^*({\bf r}_{\perp})
 \chi_m({\bf r}_{\perp}^\prime)
 \rho_{nm}(z,z^\prime,t) \; .
 \label{expansion}
\end{equation}
Here, the channel index $n$ labels the transverse states of the
trapping potential. The
transverse wavefunctions $\chi_n({\bf r}_{\perp})$ are chosen mutually
orthogonal and are eigenfunctions of the transverse part of the Hamiltonian in
Eq.~(\ref{ham}) in the sense that
\begin{equation}
 \left[-\frac{\hbar^2}{2m}\nabla^2_{{\bf r}_\perp} +
 V_t({\bf r}_\perp)\right]\chi_n({\bf r}_\perp) =
 E_n\chi_n({\bf r}_\perp) \; .
\end{equation} 
The decomposition (\ref{expansion}) in transverse and longitudinal components
of the density matrix is now inserted in the von Neumann equation,
Eq.~(\ref{3vN}). Making use of the orthogonality and the completeness of the
transverse states $\chi_n({\bf r}_{\perp})$ we obtain a one-dimensional
equation for the evolution of the density matrix:
\begin{eqnarray}
&&\left[i\hbar\frac{\partial}{\partial t} +
\frac{\hbar^2}{2m}\left(\frac{d^2}{dz^2} -
\frac{d^2}{dz^{\prime^2}}\right) -\Delta E_{lk}\right]
\rho_{lk}(z,z^\prime,t) \label{1vN}\\
&& = \sum_{n} \left[
\delta S_{ln}(z,t)\rho_{nk}(z,z^\prime,t) - \delta
S_{nk}(z^\prime,t)\rho_{ln}(z,z^\prime,t)\right]
\;.\nonumber
\end{eqnarray}
Here the abbreviation $\Delta E_{lk} = E_l - E_k$ was introduced for
the difference of transverse energy levels, and the fluctuations
$\delta S_{nk}(z)$ are defined as
\begin{equation}
 \delta S_{nk}(z,t) = \int
 d{\bf r}_{\perp}\;\chi_n({\bf r}_{\perp}) \delta
 V({\bf x},t) \chi_k^*({\bf r}_{\perp}) \label{defS} \; .
\end{equation}

The left-hand side of the reduced von Neumann equation,
Eq.~(\ref{1vN}), describes the evolution of an atom wavefunction in
the non-fluctuating trapping potential and we will hereafter
abbreviate this part by $(i\hbar\partial_t - \hat{H}_{lk})$. The terms
on the right-hand side of Eq.~(\ref{1vN}) contain the influence of the
potential fluctuations.

The fluctuations of the confining potential
are described by the matrix elements $\delta S_{nk}(z,t)$ which
imply transitions between different discrete transverse energy levels
induced by the fluctuating potential. The influence of the
fluctuating potential onto the longitudinal motion of the atomic cloud
is included in the $z$ dependence of $\delta
S_{nk}(z,t)$.

%%%%%%%%%%%%%%%%%%%%%%%%%%%%%%%%%%%%%%%%%%%%%%%%%%%%%%%%%%%%%%%%%%%%%%
\subsection{\label{sec:level2} Averaged equation of motion for the density
 matrix} 

In this section we derive the equation of motion describing the evolution of
the density matrix $\langle\rho\rangle$ averaged over the potential
fluctuations \cite{Gardiner01,vanKampen}. To keep the expressions
compact we rewrite Eq.~(\ref{1vN}) as
\begin{equation}
\left(i\hbar\partial_t - \tilde{H}\right)\rho = \delta\tilde{S}\rho
\label{compressed} \; ,
\end{equation}
where the spatial coordinates ${\bf x}$, ${\bf x}^\prime$, time $t$ and all
indices have been suppressed. The components of the quantities $\tilde{H}$ and
$\delta\tilde{S}$ are defined as
\begin{eqnarray}
 \tilde{H}_{lkij} &=&
 \hat{H}_{lk}\delta_{li}\delta_{kj}\,,\\
 \delta\tilde{S}_{lkij} &=& 
 \delta S_{li}(z,t)\delta_{kj} - \delta S_{jk}(z^\prime,t)\delta_{li}\,.
 \label{deftilde}
\end{eqnarray}
The product in (\ref{compressed}) has the meaning $\tilde{A}\rho = \sum_{ij}
A_{lkij}\rho_{ij}$. The stochastic equation (\ref{compressed}) for the density
matrix is averaged over the fluctuating potential using the standard cumulant
expansion \cite{vanKampen}. Writing the time arguments again we obtain
\begin{widetext}
\begin{equation}
 \left(i\hbar\partial_t - \tilde{H}\right)\langle\rho(t)\rangle =
 \langle\delta\tilde{S}(t)\rangle \langle\rho(t)\rangle
 - \frac{i}{\hbar}\int^t_0 dt^\prime
 \langle\langle\delta\tilde{S}(t)e^{-\frac{i}{\hbar}
 \tilde{H}t^\prime}\delta\tilde{S}(t-t^\prime)\rangle\rangle
 e^{\frac{i}{\hbar}\tilde{H}t^\prime}\langle\rho(t)\rangle
 \; . \label{cum}
\end{equation}
Here, the brackets $\langle\,\cdot\,\rangle$ denote the averaging over all
realisations of the potential fluctuations and the double brackets
$\langle\langle\,\cdot\,\rangle\rangle$ denote the second cumulant. We can
take the mean value of the potential fluctuation $\langle\delta V \rangle$ to
zero, since the static potential is already included in the Hamiltonian on the
left hand side of Eq.~(\ref{cum}) and, hence, the first term on the
right-hand side of Eq.~(\ref{cum}) vanishes.

Reinserting the explicit expressions for $\tilde{H}$, $\delta\tilde{S}$ leads
to the desired equation of motion for the density matrix
\begin{eqnarray}
 \label{finalmess}
&&\left( i\hbar\partial_t - \hat{H}_{lk}\right)
 \langle\rho_{lk}(z,z^\prime,t)\rangle= 
 -\frac{i}{\hbar}\sum_{ijmn}\int\limits^t_0 d\tau 
 \int\limits^{\infty}_{-\infty}d\tilde{z}\int\limits^{\infty}_{-\infty}
 d\tilde{z}^\prime K_{ij}(z-\tilde{z},z^\prime-\tilde{z}^\prime,t-\tau)
 \\\nonumber
 &&\times\left[\langle \delta S_{li}(z,t) 
 \delta S_{im}(\tilde{z},\tau)\rangle\delta_{kj}\delta_{jn}
 + \langle \delta S_{jk}(z^\prime,t) \delta
 S_{nj}(\tilde{z}^\prime,\tau)\rangle
 \delta_{li}\delta_{im}\right.
 \\&&\nonumber
 \left.- \langle \delta S_{li}(z,t) \delta
 S_{nj}(\tilde{z}^\prime,\tau)\rangle\delta_{im}\delta_{jk}
 -\langle \delta S_{jk}(z^\prime,t) \delta
 S_{im}(\tilde{z},\tau)\rangle\delta_{jn}\delta_{li}
 \right]\langle \rho_{mn}(\tilde{z},\tilde{z}^\prime,\tau)\rangle 
 \,.
\end{eqnarray}
\end{widetext}
The kernel $K_{ij}(z-\tilde{z},z^\prime-\tilde{z}^\prime,t)$ is the 
Fourier transform of 
\begin{equation}
K_{ij}(q,q^\prime,t) = \exp\left(-i\frac{\hbar}{2m}
\left(q^2 - {q^{\prime}}^2 \right)t -\frac{i}{\hbar}
\Delta E_{ij}t\right)\label{kernel} \;,
\end{equation}
which can be explicitly evaluated, but this has no advantage for our further
discussion.

Equation~(\ref{finalmess}) is the main result of this section. It
describes the evolution of the density matrix in a quasi-1d waveguide
under the influence of an external noise source. It is valid for an
arbitrary form of the transverse confining potential, thus, in
particular, for single and double-wire traps. The main input is the
external noise correlator and its effect on the trapped atoms. The
noise correlator depends on the concrete wire configuration. Below we
will derive a simplified form of the equation of motion
(\ref{finalmess}) under the assumption that the time scale
characterizing the fluctuations is much shorter than the time scales
of the atomic motion.

%%%%%%%%%%%%%%%%%%%%%%%%%%%%%%%%%%%%%%%%%%%%%%%%%%%%%%%%%%%%%%%%%%%%%%%%%%%%
\subsection{Noise correlation function}
\label{sec:noise}

We will now derive the noise correlator for our specific system and use
Eq.~(\ref{finalmess}) to study the dynamics. As we are interested in the
coherence of atoms in an atom-chip trap, we will consider the current noise
in the wires as the decoherence source. Fluctuations of the magnetic bias
fields $B^{(x)}_{\textrm{bias}}$ and $B^{(z)}_{\textrm{bias}}$, needed to form
the trapping potential, will be neglected.

Using the approximation of a 1d wire, the fluctuating current
density can be written as
\begin{equation}
 {\bf j}({\bf x},t)=I(z,t) \delta (x) \delta (y) \hat{{\bf z}}
 \label{currentdensity}\;,
\end{equation}
where $\hat{\bf z}$ is the unit vector in $\hat{z}$-direction. The
fluctuations of the current density are defined by
$\delta{\bf j}({\bf x},t) \equiv {\bf j}({\bf x},t) - \langle
{\bf j}({\bf x})\rangle$. It is sufficient to know $\langle\delta
I(z,t)\delta I(z^\prime,t^\prime)\rangle$ to obtain the full
current density correlation function
$\langle\delta{\bf j}({\bf x})\delta
{\bf j}({\bf x}^\prime)\rangle$. Note, that the average currents
are already included in the static potential $V_{t}(\bf r_\perp)$.

The restriction of ${\bf j}$ to the ${\bf \hat{z}}$ direction is a
reasonable assumption since we consider micro-structured wires with
small cross-section $A=l_wl_h$, i.e. wires with widths $l_w$ and
heights $l_h$ much smaller than the trap-to-wire distance $r_0$.
Transversal current fluctuations lead to surface charging and thus to
an electrical field which points in opposite direction to the current
fluctuation. This surface charging effect will suppress 
fluctuations which are slow compared to
$\omega_{RC}=\sigma\Lambda/\epsilon_0 \l_w$.
 Here, $\sigma$ is the
conductivity of the wire, $\epsilon_0$ the (vacuum) 
dielectric constant, and $\Lambda\approx
1\mbox{\AA}$ is the screening length in the metal. This leads to
RC-frequencies of $\omega_{RC}\approx 10^{13}{\mbox{Hz}}$ for wire
widths of $l_w =10\mu\mbox{m}$ and typical values for the conductivity
in a metal.
 
The characteristic time scale for the atomic motion in the trap is
given by the frequency of the trapping potential $\omega\approx
10^4\mbox{Hz}$.  Thus, considering atomic traps with
$\omega\ll\omega_{RC}$ the current fluctuations can be taken along
${\bf \hat{z}}$ as a direct consequence of the
quasi-one-dimensionality of the wire.

The current fluctuations are spatially uncorrelated as they have their
origin in electron scattering processes.
Hence, the correlator
$\langle\delta I(z)\delta I(z^\prime)\rangle$ has the form
\cite{deJong01,Blanter01}
\begin{equation}
 \langle\delta I(z,t)\delta I(z^\prime,t^\prime)\rangle =
 4k_{\textrm{B}}T_{\textrm{eff}}(z)\sigma A\,\delta(z-z^\prime) 
 \delta_{c}(t-t^\prime) 
 \label{nyquist}\;.
\end{equation} 
Here, $k_{\textrm{B}}$ is the
Boltzmann constant.  The effective
noise temperature is given by \cite{deJong01,Blanter01}
\begin{equation}
 T_{\textrm{eff}}(z)=\int dE f(E,z)[1-f(E,z)]\;,
\end{equation}
where $f(E,z)$ is the energy- and space-dependent non-equilibrium distribution
function. A finite voltage across the wire induces a change in the velocity
distribution of the electrons and the electrons are thus no longer in thermal
equilibrium. Nevertheless, the deviation from an equilibrium distribution is
small at room temperature due to the large number of inelastic scattering
processes. The effective temperature $T_{\mathrm{eff}}$ accounts for possible
non-equilibrium effects such as shot noise. However, contributions of
non-equilibrium effects to the noise strongly depend on the length $L$ of the
wire compared to the characteristic inelastic scattering lengths. E.g. strong
electron-phonon scattering leads to an energy exchange between the lattice and
the electrons. The non-equilibrium distribution is 'cooled' to an equilibrium
distribution at the phonon temperature. Thus, non-equilibrium noise sources,
such as shot noise, are strongly suppressed for wires much longer than the
electron-phonon scattering length $l_{\mathrm{ep}}$ and the noise in the wire
is essentially given by the equilibrium Nyquist noise
\cite{deJong01,Nagaev01,Kozub01}. As the wire lengths used in present
experiments are much longer than $l_{\mathrm{ep}}$, we 
$T_{\mathrm{eff}} \approx 300{\mathrm{K}}$ in all our calculations.

Finally, 
\begin{equation}
\label{eq:tauc}
\delta_{c}(t) = \frac{1}{\pi}\frac{\tau_c}{t^2 +\tau^2_c}
\end{equation}
is a representation of the delta function. 
The correlation time $\tau_c$ is given by the time scale of the electronic
scattering processes. 

\subsection{Simplified equation of motion}

The dominating source of the current noise in the wires is due to the
scattering of electrons with phonons, electrons and impurities. These
scattering events are correlated on a time scale much shorter than the
characteristic time scales of the atomic system. 
This separation of time
scales allows us to simplify the equation of motion (\ref{finalmess}).

As a consequence of Eq.~(\ref{nyquist}), the correlation
function of the fluctuating potential will be of the form
\begin{equation}
\langle \delta V({\bf x},t)\delta
V({\bf x}^\prime,t^\prime)\rangle = \delta_{c}(t-t^\prime)
\langle \delta V({\bf x})\delta
V({\bf x}^\prime)\rangle \;.
\end{equation}
Using Eq.~(\ref{defS}) we obtain
\begin{equation}
 \langle \delta S_{im}(z,t)\delta
 S_{nj}(z^\prime,t^\prime)\rangle = \delta_{c}(t-t^\prime)
 \langle \delta S_{im}(z)\delta S_{nj}(z^\prime)\rangle
 \label{corrSS}
\end{equation}
for the fluctuations of the projected potential. 

We replace all the correlation functions in the equation of motion
(\ref{finalmess}) by the expression (\ref{corrSS}). The time
integration can be performed using the fact that the averaged density
matrix $\langle\rho(\tau)\rangle$ varies slowly on the correlation
time scale $\tau_c$. Thus, $\langle\rho(\tau)\rangle$ can be evaluated
at time $t$ and taken out of the time integral. Finally, taking
$\tau_c$ to zero, the Fourier transform of the kernel
(\ref{kernel}) leads to a product of delta functions in the spatial
coordinates. Performing the remaining spatial integrations over
$\tilde z$ and $\tilde z^\prime$, Eq.~(\ref{finalmess}) reduces to
\begin{widetext}
\begin{eqnarray}
 &&\left( i\hbar\partial_t - \hat{H}_{lk}\right)
 \langle\rho_{lk}(z,z^\prime,t)\rangle = \label{final2}
 -\frac{i}{2\hbar}\sum_{mn} \left[\langle \delta S_{ln}(z) \delta
 S_{nm}(z)\rangle\langle\rho_{mk}(z,z^\prime,t)\rangle\right.\\\nonumber
 &&+ \langle \delta S_{mk}(z^\prime)\delta S_{nm}(z^\prime)
 \rangle\langle\rho_{ln}(z,z^\prime,t)\rangle
 \left.- 2\langle\delta S_{lm}(z) \delta S_{nk}(z^\prime)
 \rangle\langle\rho_{mn}(z,z^\prime,t) \rangle\right] \,.
\end{eqnarray}
\end{widetext}

The trapping potential has $N$ minima, i.e.,
the channel index $n$ can be written as $n=(\alpha,n_x,n_y)$,
$\alpha=1,...,N$. We will now assume that the minima are
well-separated in $\hat{x}\hat{y}$-direction such that we can neglect 
all matrix elements $\delta S_{ij}$ with different trap labels
$\alpha$. Under this assumption, the first and second term on the
right-hand side of Eq.~(\ref{final2}) depend only on a single trap
label. In the third term, the trap labels may be different for $\delta
S_{lm}$ as compared to $\delta S_{nk}$. Since current noise in
one particular wire generates fluctuations of the magnetic field in all
trapping wells, there will be a correlation between the fluctuations in
different traps. It is hence this third term in Eq.~(\ref{final2}) which
describes the correlation between potential fluctuations in different traps.

Using Eq.~(\ref{final2}) it is now possible to describe decoherence
induced by current noise in trapping geometries of one, two or more
parallel wires \cite{Reichel02,Folman01}. In the following sections we
will discuss decoherence in two specific trapping configurations: the
single-wire trap in Section~\ref{sec2a}, and the double-wire trap in
Section~\ref{sec2b}, which is of particular interest for atom
interferometry experiments \cite{Hinds01,Andersson01,Haensel03}.

%%%%%%%%%%%%%%%%%%%%%%%%%%%%%%%%%%%%%%%%%%%%%%%%%%%%%%%%%%%%%%%%%%%%%%
\section{\label{sec2} Trapping Geometries}

We will now consider two specific trap configurations, the single-wire trap
and the double-wire trap. The dynamics of the noise-averaged density matrix 
will be discussed using Eq.~(\ref{final2})
which was derived in the previous section. We assume that the wire
generating the potential fluctuations is one-dimensional. This assumption is
reasonable for distances $r_0$ of the trap minimum to the wire much larger
than the wire width $l_w$ and wire height $l_h$. Typical length scales are
$r_0\approx~10\mu$m to $1$mm and wire widths and heights $l_w\approx 10\mu$m
to $50\mu$m, $l_h\approx 1\mu$m \cite{Folman01}. 

To keep the notation short, we will suppress the brackets denoting the
averaging. Thus, from now on, the density matrix $\rho_{ij}$ denotes
the average density matrix.

%%%%%%%%%%%%%%%%%%%%%%%%%%%%%%%%%%%%%%%%%%%%%%%%%%%%%%%%%%%%%%%%%%%%%%
\subsection{\label{sec2a} The single-wire trap}

We are now looking at a specific magnetic trapping field having only a
single one-dimensional trapping well as shown in Fig.~\ref{figure2}.
The magnetic field is generated by the superposition of the magnetic field due
to the current $I$ in the single wire along the $\hat{z}$-axis and a
homogeneous bias field $ B^{(x)}_{\textrm{bias}}$ parallel to the chip surface
and perpendicular to the wire (see Fig.~\ref{figure0}). We additionally
include a homogeneous bias field $B^{(z)}_{\textrm{bias}}$ parallel to the
trapping well. This longitudinal bias field is experimentally needed to avoid
spin flips at the trap center. Thus the magnetic field has the form:
\begin{equation}
 {\bf B}({\bf x}) = \frac{\mu_0 I}{2\pi} \frac{1}{x^2 +
 y^2}\left(
 \begin{array}{c} -y \\ x \\0\end{array}\right) +
 \left(\begin{array}{c}
 B^{(x)}_{\textrm{bias}}\\0\\B^{(z)}_{\textrm{bias}}
 \end{array}\right) .
 \label{defBsw}
\end{equation}
The minimum of the trapping potential is located above the wire $x_0=0$ and
has a wire-to-trap distance of $r_0=y_0=\mu_0 I /(2\pi
B^{(x)}_{\textrm{bias}})$.
 
We have now all the ingredients needed to calculate the fluctuation-correlator
$\langle \delta S_{im}(z)\delta S_{nj}(z^\prime)\rangle$. Using
Eqs.~(\ref{defS}), (\ref{currentdensity}), and (\ref{nyquist}) we obtain
\begin{equation}
 \langle \delta S_{im}(z)\delta S_{nj}(z^\prime)\rangle =
 A_{imnj}J(z-z^\prime) \label{SSswgeneral}\;.
\end{equation}
The transition matrix elements are given by
\begin{equation}
\begin{split}
 A_{imnj}= &\quad k_{\textrm{B}}T_{\textrm{eff}}\sigma A
 \left(\frac{\mu_0 g_F \mu_B}{\pi}\frac{B^{(x)}_{\textrm{bias}}}
 {B^{(z)}_{\textrm{bias}}}\right)^2\\
&\times\int d{\bf r}_{\perp}\;\chi_i({\bf r}_{\perp})(y-y_0)
\chi_m^*({\bf r}_{\perp})\\
 &\times\int
 d{\bf r}^\prime_{\perp}\;\chi_n({\bf r}^\prime_{\perp})
 (y^\prime-y_0)\chi_j^*({\bf r}^\prime_{\perp})\label{A}\;.
\end{split}
\end{equation}
The spatial dependence of the noise correlator is given by
\begin{equation}
 J(z) = \frac{1}{r^5_0}\int\limits^\infty_{-\infty} d\tilde{z}
 \left[1+\tilde{z}^2\right]^{-\frac{3}{2}}\left[1+
 \left(z/r_0-\tilde{z}\right)^2\right]^{-\frac{3}{2}}
 \label{J}\;.
\end{equation}
A more detailed derivation of the correlation function is given in
Appendix \ref{appendixA}. The integration in Eq.~(\ref{A}) can be done
explicitly using harmonic oscillator states for $\chi_n({\bf r}_\perp)$,
which is a good approximation as long as $B^{(z)}_{\textrm{bias}}$ is
of the order of $B^{(x)}_{\textrm{bias}}$ and the wire-to-trap
distance $r_0$ is much larger than the transverse width $w$ of the
trapped state. Both requirements are usually well satisfied in
experiments. The dashed line in Fig.~\ref{figure2} shows the harmonic
 approximation to the trapping potential. Using
the harmonic approximation we can characterize the steepness of the
trapping potential by its trap frequencies. It turns out that the two
frequencies coincide,
\begin{equation}
 \omega=\sqrt{\frac{2\mu_B g_F}{m
 B^{(z)}_{\textrm{bias}}}}\frac{B^{(x)}_{\textrm{bias}}}{y_0}
\label{singleomega}\; ,
\end{equation}
i.e., the trap potential can be approximated by an isotropic 2d
harmonic oscillator (2d-HO).

 After integration we obtain
\begin{equation}
 A_{imnj}=A_0 \;\delta_{i_xm_x}\tilde{\delta}_{i_y m_y}
 \delta_{n_xj_x}\tilde{\delta}_{n_y j_y}\;,\label{SSsinglewire}
\end{equation}
where
\begin{equation}
 \tilde{\delta}_{i_ym_y} = \sqrt{m_y+1}
 \;\delta_{i_y,m_y+1}+\sqrt{m_y}\;\delta_{i_y,m_y-1}\;,\label{tildedelta}
\end{equation}
and the indices $n_x$, $n_y$ denote the energy levels in $\hat{x}$ and
$\hat{y}$ direction of the 2d-HO. The prefactor $A_0$ is
\begin{equation}
 A_0=k_{\textrm{B}}T_{\textrm{eff}}\sigma A
 \left(\frac{\mu_0g_F\mu_B}{\pi}\frac{B^{(x)}_{\textrm{bias}}}
 {B^{(z)}_{\textrm{bias}}}\right)^2\frac{w^2}{2}\label{defA0}
\end{equation}
and $w=\sqrt{\hbar/(m\omega)}$ is the oscillator length of the harmonic
potential.

Before moving on to derive the equation of motion for the averaged density
matrix, let us discuss some consequences of expression (\ref{SSsinglewire}).
Inspection of the coefficients $A_{imnj}$ in Eq.~(\ref{SSsinglewire}) shows
that there is no direct influence onto the longitudinal motion of the atom,
since the matrix element $A_{iijj}$ vanishes. This result is not unexpected,
because we consider only current fluctuations along the wire, which give
rise to fluctuations in the trapping field along the transverse directions of
the trap potential. In fact, only transitions among energy levels of the
$\hat{y}$-component of the 2d-HO give a non-vanishing contribution. An
explanation can be obtained by examining the change of the trap minimum
position under variation of the current in the wire. Changing the current by
$\delta I$ leaves the trap minimum in $\hat{x}$-direction unchanged at
$x_0=0$, but shifts the $\hat{y}$-trap minimum by $\delta y_0= \mu_0 \delta I
/(2\pi B^{(x)}_{\textrm{bias}})$ \cite{misalign}. Even though there is no
direct coupling to the motion along the wire there can still be spatial
decoherence in $\hat{z}$-direction as we will see in the final result for
$\rho$ obtained from Eq.~(\ref{final2}). However, this requires
transitions to neighbouring transverse energy levels.

Substituting Eqs.~(\ref{SSswgeneral}) and (\ref{SSsinglewire}) in
Eq.~(\ref{final2}) leads to an equation of motion for $\rho$ for the
single-wire trap configuration. Instead of discussing the general
equation of motion of $\rho$ we will only consider the case where the
two lowest energy levels of the transverse motion are taken into
account. The restriction to the lowest energy levels corresponds to
the situation of the atomic cloud being mostly in the ground state of
the trap and having negligible population of higher energy
levels. This situation is realistic, if the energy spacing of the
discrete transverse states is large compared to the kinetic energy.
Nevertheless, the result obtained from the two-level model provides a
reasonable estimate for the decoherence of an atomic cloud, even if
many transverse levels are populated. Equation~(\ref{SSsinglewire})
shows that only next-neighbour transitions are allowed. As $A_{imnj}$
is a product of two next-neighbour transitions there can only be
contributions of the next two neighbouring energy levels. Higher
energy levels do only contribute to the decoherence by successive
transitions which are of higher order in $A_{imnj}$ and hence are
negligible.

\begin{figure}
\includegraphics[width=8cm]{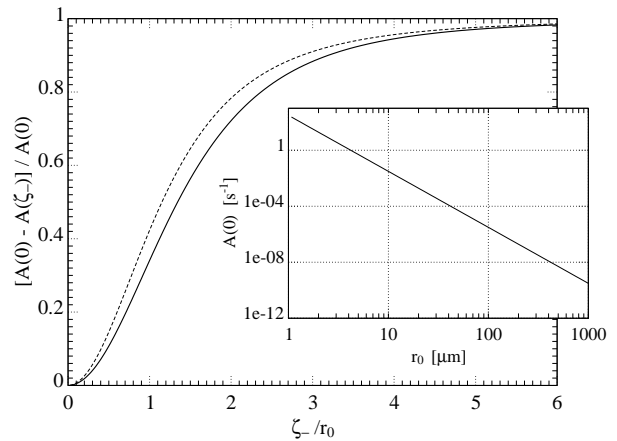}

\caption{Spatial dependence of the decay rate
 $\bar{\Gamma}_{\mathrm{dec}}(\zeta_-)=A(0)-A(\zeta_-)$ for the diagonal
 elements $\bar{\rho}_{ii}(\zeta_-,t)$, Eq.~(\ref{averagedecay}). The 
 wire-to-trap distance is $r_0=100\mu\textrm{m}$ and the trap frequency is
 $\omega\approx 2\pi\times 10 \textrm{kHz}$. The dashed line is the
 approximation Eq.~(\ref{approxJ}) for $A(\zeta_-)$. Inset: Decay rate
 $A(0)$ as function of $r_0$. The parameters chosen in the plots correspond
 to $^{87}\textrm{Rb}$ trapped at $T_{\textrm{eff}}=300\textrm{K}$ in a
 magnetic trap with a gold wire of conductivity 
 $\sigma_{\textrm{Au}}~=~4.54\cdot 10^7 \Omega^{-1}\textrm{m}^{-1}$ and a
 cross-section of $A=2.5\mu{\textrm{m}} \times 5\mu{\textrm{m}}$. The bias
 fields chosen are $B^{(x)}_{\textrm{bias}}=80\textrm{G}$ and
 $B^{(z)}_{\textrm{bias}}=2\textrm{G}$. \label{figure3}}
\end{figure}

We rewrite Eq.~(\ref{final2}) for the subspace of the two lowest eigenstates
in a matrix equation for the density vector
\begin{equation}
{\bf{\rho}} = \left(\rho_{00},
\rho_{11},\rho_{10},\rho_{01}\right)\label{densityvector}\;.
\end{equation}
The indices of the averaged density matrix $\rho_{lk}$ denote the
transverse state $l=(l_x,l_y)$ but we are only writing the $l_y$
component of the label as $\rho$ can only couple to states with the
same $x$-state (i.e. only transitions between different $y$-states of
the 2d-HO are allowed). The $x$-label is $l_x = 0$ for all
states under consideration. Hence Eq.~(\ref{final2}) can be written
as a matrix equation for the density matrix vector
Eq.~(\ref{densityvector})
\begin{equation}
 \left[i\hbar\partial_t -\tilde{H}(z,z^\prime)\right]{\bf{\rho}}
(z,z^\prime,t) = 
 -i\hbar\tilde{A}(z-z^\prime){\bf{\rho}}(z,z^\prime,t)\,.
 \label{master}
\end{equation}
The matrix $\tilde{A}$ is defined as
\begin{equation}
 \tilde{A}(\zeta_-)=\left(\begin{array}{cccc} 
 A(0)&-A(\zeta_-)&0&0\\-A(\zeta_-)&
 A(0)&0&0\\0&0&A(0)&-A(\zeta_-)\\0&0&-A(\zeta_-)&A(0)\end{array}\right)\,,
\end{equation}

where $A(\zeta_-)=\frac{1}{\hbar^2}A_0J(\zeta_-)$ and $\zeta_{-}=z-z^\prime$.
The equations for the diagonal elements, i.e., $\rho_{00}$,
$\rho_{11}$, and for the off-diagonal elements, i.e. $\rho_{01}$,
$\rho_{10}$ decouple. However, the decoupling is a consequence of the
restriction to the two lowest energy levels and is not found in
the general case. Yet, the decoupling allows to find an explicit
solution for the time evolution of the diagonal elements. The matrix
equation proves to be diagonal for the linear combinations
$\rho^{\pm}\equiv\rho_{00} \pm \rho_{11}$ and introducing the new
coordinates $\zeta_{-}=z-z^\prime$ ,
$\zeta_{+}=\frac{1}{2}(z+z^\prime)$ we obtain the general solution
\begin{equation}
 \rho^{\pm}_k(\zeta_-,\zeta_+,t)=
 R^{\pm}_k(\zeta_- - \frac{\hbar k}{m}t) e^{ik\zeta_+} 
 e^{-\Gamma^{\pm}_k(\zeta_-,t)t} \label{rhopm}\;,
\end{equation}
where the decay is described by
\begin{equation}
 \Gamma^{\pm}_k(\zeta_-,t)=A(0)
 \mp\frac{1}{t}\int\limits_0^t dt^\prime A(\zeta_- -\frac{\hbar
 k}{m}t^\prime)\,.\label{singledecay}
\end{equation}
The function $R^{\pm}_k$ is fixed by the initial conditions, i.e. by
the density matrix at time $t=0$ :
\begin{equation}
 R^{\pm}_k(\zeta_-) = \int\limits_{-\infty}^\infty d\zeta_+
 e^{-ik\zeta_+}\rho^{\pm}(\zeta_-,\zeta_+,t=0) \,.
\end{equation}
Note that $\Gamma$ in Eq.~(\ref{singledecay}) is a function of the
spatial variable $\zeta_-$ and the wavevector $k$, and is in general
not linear in the time argument $t$. Adding and subtracting the
contributions $\rho^\pm$ finally leads to the following expressions
for the diagonal elements of the density matrix
\begin{equation}
\rho_{00}(\zeta_-,\zeta_+,t)=\frac{1}{2}\int\limits_{-\infty}^\infty 
\frac{dk}{2\pi}[
\rho^+_k(\zeta_-,\zeta_+,t) + \rho^-_k(\zeta_-,\zeta_+,t)]\,.
\label{rhofinal}
\end{equation}
The result for $\rho_{11}$ can be obtained from (\ref{rhofinal}) by
replacing the plus sign between the terms by a minus sign. 
To get a feeling for the spatial correlations we trace
out the center of mass coordinate $\zeta_+$ :
\begin{eqnarray}
\bar{\rho}_{00}(\zeta_-,t)&=&\int_{-\infty}^\infty
 d\zeta_+\;\rho_{00}(\zeta_-,\zeta_+,t)\nonumber\\ &=&
 e^{-\left[A(0)-A(\zeta_-)\right]t}\nonumber\\
 &\times&\frac{1}{2}\left[\bar{\rho}_{00}(\zeta_-,0)
\left(1+e^{-2A(\zeta_-)t}\right)\right.\nonumber\\
 &+&\left. \bar{\rho}_{11}(\zeta_-,0)
\left(1-e^{-2A(\zeta_-)t}\right)\right]\label{averagedecay}
 \;.
\end{eqnarray}
Equation~(\ref{averagedecay}) shows that we can distinguish two decay
mechanisms. There is an overall decay of the spatial off-diagonal
elements with a rate $\bar{\Gamma}_{\textrm{dec}}(\zeta_-) =
[A(0)-A(\zeta_-)]$. This decay only affects the density matrix for
$\zeta_-\neq 0$, thus suppressing the spatial coherence. The spatial
correlation of the potential fluctuations can be read off
Fig.~\ref{figure3}, showing $\bar{\Gamma}_{\textrm{dec}}(\zeta_-)$.
We find for the potential fluctuations a correlation length $\xi_c$ of
the order of the trap-to-chip surface distance $\xi_c\approx r_0$. The
correlation length $\xi_c$ must not be confused with the coherence
length describing the distance over which the transport of an atom
along the trapping well is coherent. This coherence length is
described by the decoherence time and the speed of the moving
wavepacket. The correlation length $\xi_c$ characterizes the distance
over which the potential fluctuations are correlated.

\begin{figure}[t]
\includegraphics[width=7cm]{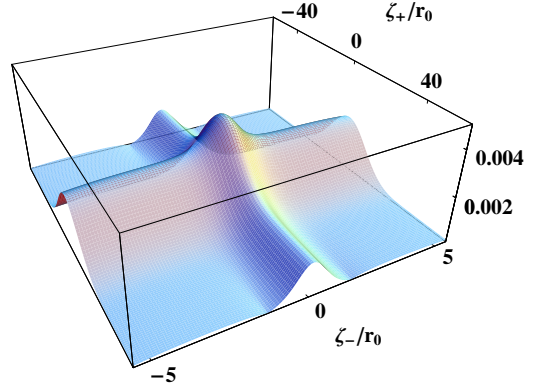}
\caption{Time evolution of $|\rho_{00}(\zeta_-,\zeta_+,t)|$ after $t=2/A(0)$.
 At $t=0$ the wavepacket is a
 Gaussian wavepacket of spatial extent $w_z=20r_0$. The 
wire-to-trap distance is $r_0=5\mu\textrm{m}$ and all 
other parameters correspond
to those used in Fig.~\ref{figure3}. 
$A(0)\approx 0.5{\mathrm{s}}^{-1}$ is given by
 Eq.~(\ref{finalsingledecay}). The density 
matrix shows damped oscillations which become more pronounced as $\zeta_+$ 
increases. 
\label{figure4a}}
\end{figure}

\begin{figure}[t]
\includegraphics[width=8.5cm]{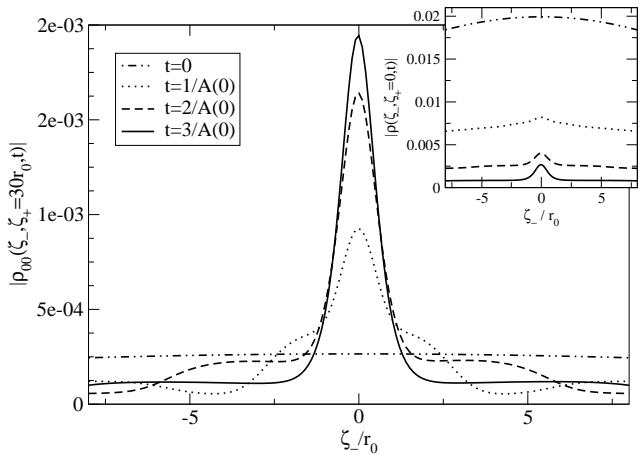}
\caption{Decay of the absolute value $|\rho_{00}(\zeta_-,\zeta_+=30r_0,t)|$
 and $|\rho_{00}(\zeta_-,\zeta_+=0,t)|$ (inset). The wave packet is a
 Gaussian wavepacket of spatial extent $w_z=20 r_0$ at $t=0$. The 
wire-to-trap distance is $r_0=5\mu\textrm{m}$. All other parameters are the 
same as used in Fig.~\ref{figure3}. $A(0)\approx 0.5 {\mathrm{s}}^{-1}$ is given by 
Eq.~(\ref{finalsingledecay}). The dashed line ($t=2/A(0)$) 
 corresponds to cuts of Fig.~\ref{figure4a} along $\zeta_-$ for $\zeta+=30 r_0$
 (inset $\zeta_+=0$). Spatial correlations of the
 potential fluctuations are restricted to a narrow band around $\zeta_- =0$.
 The width of this band is on the order of $r_0$.
 The density matrix shows damped 
oscillations which become  more pronounced with increasing $\zeta_+$.
 The increase 
of $|\rho_{00}(\zeta_-,\zeta_+=30r_0,t)|$ for increasing time $t$
 is a consequence of the spreading of the wavepacket. 
\label{figure4b}}
\end{figure}

The second mechanism describes the equilibration of excited and ground
state, quantified by the diagonal elements $\bar{\rho}_{00}(0,t)$,
which occurs at a rate $\bar{\Gamma}_{\textrm{pop}}(0)=2A(0)$. Here
equilibration means that, due to this mechanism, the probability to be
in the ground state, i.e. $\rho_{00}(0,t)$, tends to 1/2. Of course,
at the same time the probability $\rho_{11}(0,t)$ to be in the excited
state approaches 1/2 at the same rate. Note, that for the spatial
off-diagonal matrix elements the equilibration rate depends on
$\zeta_-$.

We will now take a closer look onto the quantity $A(0)$ for realistic
trap parameters. Using $J(0)=\frac{3\pi}{8y_0^5}$ and
Eq.~(\ref{singleomega}), Eq.~(\ref{defA0}) leads to
\begin{equation}
 A(0)= \frac{3\pi}{2} k_{\textrm{B}}T_{\textrm{eff}}\sigma
 A\frac{B^{(x)}_{\textrm{bias}}}{2\hbar
 \sqrt{m}}\left(\frac{\mu_0}{4\pi}\right)^2\left(\frac{2\mu_B
 g_F}{B^{(z)}_{\textrm{bias}}}
 \right)^{\frac{3}{2}}\frac{1}{r_0^4}\label{finalsingledecay}\;.
\end{equation} 
The inset of Fig.~\ref{figure3} plots $A(0)$ over $r_0$ for reasonable
trap parameters. Equation (\ref{finalsingledecay}) shows that the
decay rate is scaling on the wire-to-trap distance as
$1/r_0^4$ giving a rapid increase of decoherence effects once the
atomic cloud is brought close to the wire.

As specific example we want to study the time evolution of the full
density matrix in the ground state. Figure~\ref{figure4a} and 
Fig.~\ref{figure4b} show the time
evolution of the absolute value of the density matrix element
$|\rho_{00}(\zeta_-,\zeta_+,t)|$ for a Gaussian wave packet of spatial
extent $w_z = 20r_0$ and a wire-to-trap distance of
$r_0=5\mu\textrm{m}$. Initially, at time $t=0$ all other elements of
the density matrix are zero. The time evolution is calculated for the
reduced subspace using Eq.~(\ref{rhopm}) and Eq.~(\ref{rhofinal}). For
the spatial correlation we use the approximation (dashed line in
Fig.~\ref{figure3})
\begin{equation}
 A(\zeta_-)\approx 4\frac{A_0}{\hbar^2 r_0^5}
 \left[\left(\frac{32}{3\pi}\right)^{\frac{2}{3}} + 
 \left(\frac{\zeta_-}{r_0} \right)^2\right]^{-\frac{3}{2}}
 \label{approxJ}\,.
\end{equation}
The time evolution of the Gaussian wavepacket shows a strip of
$|\zeta_-| < l_c\approx r_0$ in which the density matrix decays much
slower. This is a consequence of the $\zeta_-$ dependence of the
decay rate shown in Fig.~\ref{figure3}. The spatial correlation length
of the wavepacket can hence be read of Fig.~(\ref{figure4b}) as
$l_c\approx r_0$. In addition, a damped oscillation in the relative
coordinate $\zeta_-$ is arising which is getting more pronounced for
larger values of $\zeta_+$. Figure~\ref{figure4b} shows cuts along the
$\zeta_-$-direction for different values of $\zeta_+$. The origin of
the damped oscillations is the $k$-dependence of the equilibration and
decoherence mechanism described by $\Gamma^{\pm}_{k}$ in
Eq.~(\ref{singledecay}). To demonstrate the influence of the
$k$-dependent damping we assume that $\Gamma^{\pm}_k = A(0)\mp [
A(\zeta_-) +\beta(\zeta_-,t) k]$. Choosing $\Gamma$ linear in $k$
leads to a modulation of the density matrix by a factor proportional
to $\cosh(i\alpha_0\beta + \alpha_1)$, describing damped
oscillations. The $\alpha_{0/1}$ are real functions of $\zeta_{\pm}$
and $t$. The oscillations arise only for non-vanishing $\beta$. Decay
rates which are $k$-independent do not show oscillations.

The decay described by $\Gamma^\pm_k$ in Eq.~(\ref{singledecay}) is
however not linear, but includes higher powers in $k$. Hence the
simple linear model $\Gamma^{\pm}_k = A(0)\mp [ A(\zeta_-)
+\beta(\zeta_-,t) k]$ describes the oscillations in
Fig.~\ref{figure4a} and Fig.~\ref{figure4b} only qualitatively.

\begin{figure}[tbh]
\includegraphics[width=8.5cm]{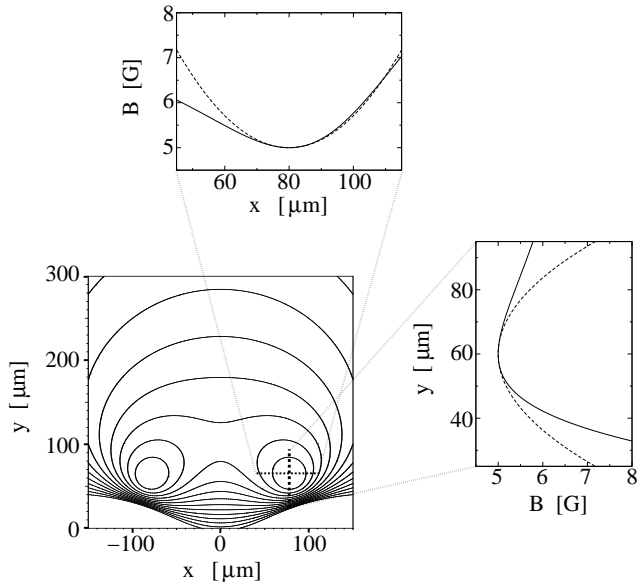}
\caption{Contour-plot of the magnetic field in the double-wire
 geometry for $B^{(x)}_{\mathrm{bias}}=
 10{\mathrm G}$, $B^{(z)}_{\mathrm{bias}}= 5{\mathrm G}$, and
 $I=0.3{\mathrm A}$. The wires are located at $x=\pm
 100\mu$m and $y=0$. Two potential minima are located roughly above the
 current-carrying wires. The upper (right) plot shows a horizontal
 (vertical) cut through the right potential minimum. The dashed lines
 in the upper and the right plot show the harmonic approximation to
 the trapping field.\label{figure6}}
\end{figure}

%%%%%%%%%%%%%%%%%%%%%%%%%%%%%%%%%%%%%%%%%%%%%%%%%%%%%%%%%%%%%%%%%%%%%%
\subsection{\label{sec2b} The double-wire trap}

The second configuration which we discuss is a double-wire trap
\cite{Hinds01}. A system of two parallel trapping wells is of special
interest for high-precision interferometry
\cite{Hinds01,Andersson01,Haensel03} or beam-splitter geometries
\cite{Andersson02}. The effect of decoherence is one of the key issues
in these experiments.

Form and characteristic of the double-wire trap is described in
\cite{Hinds01,Folman01} so that we will only briefly introduce the
main features of the double-wire trapping potential.
Figure~\ref{figure0}b shows the setup for the trapping field, which is
generated by two infinite wires running parallel to the
$\hat{z}$-axis, separated by a distance $d$, and a superposed
homogeneous bias field $B^{(x)}_{\textrm{bias}}$ perpendicular to the
current direction. A second bias field pointing along the $\hat{z}$-direction,
 $B^{(z)}_{\textrm{bias}}$, is added in experimental setups to
avoid spin flips. Thus the trapping field is
\begin{widetext}
\begin{equation}
 {\bf B}({\bf x}) = \frac{\mu_0 I}{2\pi} \left[ \frac{1}{\left(
 x+\frac{d}{2}\right)^2 +y^2}\left(\begin{array}{c}
 -y\\x+\frac{d}{2}\\0\\ \end{array}\right) +
 \frac{1}{\left(x-\frac{d}{2}\right)^2 +y^2} \left(\begin{array}{c}
 -y\\x-\frac{d}{2}\\0\end{array}\right)\right] + 
 \left(\begin{array}{c}
 B^{(x)}_{\textrm{bias}}\\0\\B^{(z)}_{\textrm{bias}}
\end{array}\right) \,.\label{defBdw}
\end{equation}
\end{widetext}
There are two different regimes for the positions of the trap-minima.
Defining a critical wire separation $\bar{y}_0~=~\mu_0 I/(2\pi
B^{(x)}_{\textrm{bias}})$ we can distinguish the situation of having
$d>2\bar{y}_0$ where the trap minima are located on a horizontal line
at $x^{L/R}_0 = \mp\sqrt{d^2/4-\bar{y}_0^2}$ and $y_0=\bar{y}_0$, and
the situation of $d<2\bar{y}_0$ where the trap minima are positioned
on a vertical line at $x_0=0$ and
$y^{\pm}_0=\bar{y}_0\pm\sqrt{\bar{y}_0^2 - d^2/4}$. The two trap
minima overlap and form a single minimum for $d=2\bar{y}_0$.

Further on we will restrict ourselves to the regime $d>2\bar{y}_0$
since in this configuration we have two horizontally spaced, but
otherwise identical trap minima. The two trap minima form
 a pair of parallel waveguides.
This kind of geometry has been suggested for an interference device
\cite{Hinds01,Andersson01}. Figure~\ref{figure6} shows a contour-plot
of the double trap in the $d>2\bar{y}_0$ regime.

We now analyze the double-wire setup along the lines of Section \ref{sec2a}. 
Firstly, an explicit expression for the correlation function
$\langle \delta S_{im}(z)\delta S_{nj}(z^\prime)\rangle$ is
needed. After some calculation we obtain
\begin{widetext}
\begin{eqnarray}
 &&\langle \delta S_{im}(z)\delta S_{nj}(z^\prime)\rangle
  =A_0\frac{8}{w^2}\frac{ x_{0}^{\alpha}
 x_{0}^{\beta}}{d^2} \label{SSgeneraldw}
 \sum_{\gamma=L,R} J^\gamma_{\alpha\beta}(z-z^\prime)
 \int d{\bf r}_{\perp}
 \int d{\bf r}_{\perp}^\prime\\\nonumber&&
 \times\chi_i({\bf r}_{\perp})\left[(y - y_0) - 
 \frac{y_0(x - x_{0}^{\alpha})}{(x_{0}^{\alpha}
 +\epsilon_\gamma\frac{d}{2})}\right]\chi^*_m({\bf r}_{\perp})\,
 \chi_n({\bf r}_{\perp}^\prime)\left[ (y^\prime - y_0) -
 \frac{y_0(x^\prime - x_{0}^{\beta})}{(x_{0}^{\beta}
 +\epsilon_\gamma\frac{d}{2})}\right]\chi^*_j({\bf r}_{\perp}^\prime)\,,
\end{eqnarray}
where the spatial correlation is now given by
\begin{equation}
 J^{\gamma}_{\alpha\beta}(z-z^\prime)=\int\limits^\infty_{-\infty}
 d\tilde{z}\left[\left(x_0^\alpha -
 \epsilon_\gamma\frac{d}{2}\right)^2 
 + y_0^2 +
 \tilde{z}^2\right]^{-\frac{3}{2}}
 \left[\left(x_0^\beta - \epsilon_{\gamma}\frac{d}{2}\right)^2 + y_0^{2} +
 (z-z^\prime
 -\tilde{z})^2\right]^{-\frac{3}{2}}\label{defJ}\;.
\end{equation}
\end{widetext}
Here, $\alpha$ is the trap label corresponding to $i$ and $m$, and
$\beta$ the trap label corresponding to $n$ and $j$, the indices which
occur in $\langle \delta S_{im}(z)\delta S_{nj}(z^\prime)\rangle$. The
wires are located at $x=d_{L/R}=\mp d/2$ and $\epsilon_\gamma$ is
defined as $\epsilon_L = -1$, $\epsilon_R = 1$. The result
Eq.~(\ref{SSgeneraldw}) uses the assumption that the extent of the
wavefunction $w$ is much smaller than $y_0$ and much smaller than the
trap minimum offset $|x^{L/R}_0|$ from the $\hat{x}$-axis, i.e., $w\ll
|x^{L/R}_0|$ and $w\ll y_0$. This approximation breaks down if the
wire separation $d$ approaches the critical separation
distance $2\bar{y}_0$ at which the two trap minima merge.

\begin{figure}[tbh]
\includegraphics[width=8.5cm]{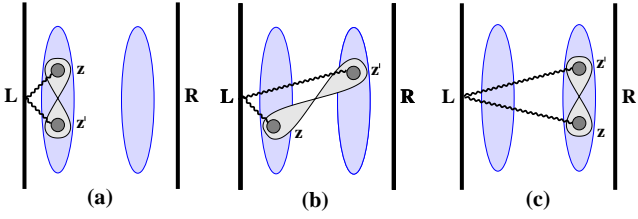}
\label{part}
\caption{Contributions of the noise in the {\it{left}} wire to the
 decoherence of $\rho_{ij}(z,z^\prime,t)$. The figures
 show only the coupling of one point on the wire to the atomic state.
 To get the total contribution to the decoherence an
 integration over all wire elements coupling to the atomic state has
 to be performed.
 \label{figure5}}
\end{figure}

Instead of going into details of the derivation we will discuss
the different contributions qualitatively. As the system is invariant
under mirror imaging at the $\hat{y}$-$\hat{z}$-plane, corresponding
to an invariance under interchange of the left (L) trap and right (R)
trap label, only the following three contributions have to be
distinguished (see Fig.~\ref{figure5}).

\begin{table}
\begin{tabular}{|c||c|c|c|c|}
\hline& $J^L_{LL}(\zeta_-)$& $J^L_{LR}(\zeta_-)$ 
& $J^L_{RR}(\zeta_-)$\\
\hline\hline $\zeta_-=0$&$ \frac{1}{\bar{y}_0^5}$
& $\frac{2}{\bar{y}_0^2d^3}$&$\frac{1}{d^5}$\\
\hline $\zeta_- \ll\bar{y}_0$& $ \frac{1}{\bar{y}_0^5}$
&$\frac{2}{d^3\bar{y}_0^2}\left(1-\frac{3}{2}\frac{\bar{y}_0^2}{d^2}\right)$
& $ \frac{1}{d^5}$\\
\hline $\bar{y}_0 \ll \zeta_- \ll d$ & $\frac{4}{\bar{y}_0^2\zeta_-^3}$
&$\frac{2}{d^3\bar{y}_0^2}\left(1-\frac{3}{2}\frac{\zeta_-^2}{d^2}\right)$
&$ \frac{1}{d^5}$\\
\hline $ d\ll \zeta_-$&$\frac{4}{\bar{y}_0^2\zeta_-^3}$
&$\frac{2}{\zeta_-^3}\left(\frac{1}{\bar{y}_0^2}+\frac{1}{d^2}\right)$
&$\frac{4}{d^2\zeta_-^3}$\\\hline
\end{tabular}
\caption{Estimate of the contributions to the decoherence. The columns for 
 $J^L_{LL}$, $J^L_{LR}$, and $J^L_{RR}$ correspond to contributions (a), (b)
 and (c) of Fig.~\ref{figure5} respectively. 
The $J^{\gamma}_{\alpha\beta}(\zeta_-)$ 
are given by Eq.~(\ref{defJ}) as a function of 
$\zeta_-= z-z^\prime$. The approximation assumes $\bar{y}_0 \ll d$.
\label{table1}}
\end{table}

\begin{itemize}
\item[(a)] The influence of current fluctuations in the 
 {\it left wire} onto an atom in a state which is localized in the
 {\it left arm} of the trapping potential (Fig.~\ref{figure5}a).
\item[(b)] The influence of current fluctuations in the {\it left
 wire} onto an atom which is in a superposition of {\it a
 state localized in the left arm} and {\it a state localized in
 the right arm} of the trapping potential (see
 Fig.~\ref{figure5}b).
\item[(c)] The influence of current fluctuations in the {\it left
 wire} onto an atom in a state which is localized in the
 {\it right arm} of the trapping potential (Fig.~\ref{figure5}c).
\end{itemize}

All other contributions are obtained by symmetry operations.
Contribution (a) is given by the term proportional to
$J^{L}_{LL}$ and is the dominating decoherence source in
the regime $d>2\bar{y}_0$, as all other contributions 
($J^L_{LR}$, $J^L_{RR}$ shown in the third and forth column of
Table~\ref{table1}) are suppressed by orders of $\bar{y}_0/d$.

The contribution in Fig.~\ref{figure5}b is of particular interest as
it describes the cross-correlations between the noise in the left and
right trapping well i.e. given by terms proportional to
$J^{L}_{RL}$ and $J^{L}_{LR}$. Current fluctuations
in one of the wires give rise to magnetic field fluctuations in all
the trapping wells. Magnetic field fluctuations are therefore not
uncorrelated. These correlations are however suppressed for
 $\bar{y}_0\ll d$ as can be seen in Table~\ref{table1}.

Contributions of type (c), given by terms proportional to
$J^{L}_{RR}$, are the smallest contributions to
decoherence as they describe the influence of the left wire current
noise onto the atomic cloud in the more distant right trap. Terms of
type (c) are thus negligible for wire distances $d\gg2\bar{y}_0$ as
the magnetic field decreases with $1/r$ and fluctuations in the
trapping field are dominated by the nearest noise source i.e.
contributions of type (a). In analogy to the proceeding Section
\ref{sec2a} we are going to approximate the transversal wavefunctions
$\chi_n$ in Eq.~(\ref{SSgeneraldw}) by 2d-HO states. However one has
to be careful as this approximation holds only if the two trapping
wells are sufficiently separated such that the mutual distortion of
magnetic field is small. If the distance between the two traps gets
close to the critical value $d=2\bar{y}_0$, a double-well structure
arises and the harmonic approximation breaks down. We estimate the
validity of the approximation by calculating the distance $d_c$ at
which the local potential maximum, separating the two trapping wells,
is of the order of the ground state energy of the isolated single
trap. The trap frequency for the double-wire configuration extracted
from the harmonic approximation is
\begin{equation}
 \omega_d = 2\omega \frac{|x_0|}{d} =2\omega
 \frac{\sqrt{d^2/4-\bar{y}_0^2}}{d} \;,
\end{equation} 
where $\omega$ is the single-wire trap frequency given in
Eq.~(\ref{singleomega}). Thus the ground state energy $E_0=\hbar\omega$ 
of the single-wire trap
corresponds to the limit of infinite separation $d=\infty$.
This energy $E_0$ is the maximum value for the ground state energy 
$\hbar\omega_d$ of the double-wire trap under variation of $d$. Hence $E_0$ 
is an upper bound for the ground state energy of the double well 
in the case $d>\bar{y}_0$.
 Using the procedure described above, we get
the condition $d > \sqrt{8}\,\bar{y}_0$, restricting the region in
which the transversal potential can be approximated as two independent
2d-HOs. The dashed line in Fig.~\ref{figure6} shows an example for the
harmonic approximation to the trapping potential of the double-wire
setup.

Performing the integration over the transverse coordinates 
${\bf r}_{\perp}$, using 2d-HO states for $\chi_n$, gives transition matrix
elements
\begin{widetext}
\begin{eqnarray}
 &&\langle \delta S_{im}(z)\delta S_{nj}(z^\prime)\rangle
 = 4A_0\frac{ x_{0}^{\alpha}x_{0}^{\beta}}{d^2}
 \sum_{\gamma} J^\gamma_{\alpha\beta}(z-z^\prime)\\\nonumber
 &&\times \; \left[\tilde{\delta}_{i_y m_y} +
 \tilde{\delta}_{i_x m_x}\frac{\epsilon_{\alpha} y_0}{(x_{0}^{\alpha}
 +\epsilon_\gamma\frac{d}{2})}\right]
 \left[\tilde{\delta}_{n_y j_y} +
 \tilde{\delta}_{n_x j_x}\frac{\epsilon_{\beta} y_0}{(x_{0}^{\beta}
 +\epsilon_\gamma\frac{d}{2})}\right]\,,\label{DW2dHO}
\end{eqnarray}
\end{widetext}
where the functions $\tilde{\delta}_{ij}$ have been defined in
Eq.~(\ref{tildedelta}). For simplicity we choose the current in the
left and right wire to be the same for the derivation of
Eq.~(\ref{DW2dHO}).

Comparing the matrix elements obtained in Eq.~(\ref{DW2dHO}) with
those in Eq.~(\ref{SSsinglewire}) for the single-wire trap, we find
the following differences. First of all it has to be noted that the
transitions are no longer restricted to the $\hat{y}$-components of
the 2d-HO. This is a direct consequence of the geometry as the
positions of the trap minima are now sensitive to current fluctuations
in $\hat{x}$- and $\hat{y}$-direction whereas the $\hat{x}$-position
for the single-wire trap minimum was independent of the current
strength in the conductor. Secondly, the prefactor of the
transition matrix element in Eq.~(\ref{DW2dHO}) is now a function of the
spatial coordinates $x_{0}^\alpha$, taking the geometric changes of the trap
minimum position under variation of the wire separation $d$ into
account.

\begin{figure}[tbh]
 \includegraphics[width=8cm]{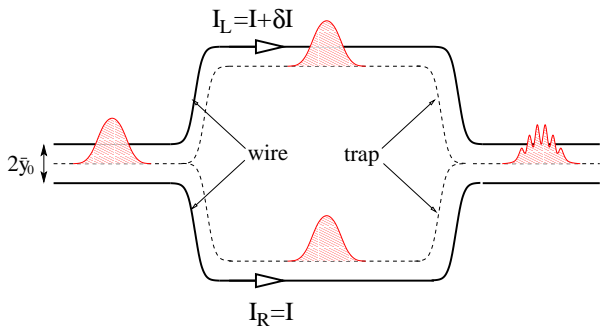}
 \caption{Schematic setup of an interferometer for cold atoms using 
 a double-wire trap. A wavepacket is coming in from the left single
 waveguide formed by a double wire trap with a wire spacing of
 $d=2\bar{y}_0$. The wavepacket is split into a coherent
 superposition of states localized in the left and right arm of the
 double waveguide.
 Inducing a potential difference, e.g. by applying a
 current difference $\delta I$ in the left and right wire, gives
 rise to a phase shift. Remerging the wavepackets results into an
 interference pattern in the atom density. For a more detailed
 description of atom
 interferometers of this type see Ref. \cite{Andersson01}.
\label{figure7}}
\end{figure}

Inserting Eq.~(\ref{DW2dHO}) in Eq.~(\ref{final2}) leads to a set of
equations for the dynamics of the averaged density matrix. For
simplicity we restrict ourselves to the ground state and first excited
state of the 2d-HO. Having in mind a spatial interferometer as
suggested in Ref. \onlinecite{Andersson01}, the most interesting
quantity is the density matrix element describing the coherent
superposition of a state in the left arm with a state in the right
arm. Figure \ref{figure7} shows the geometry of the interferometer. A
wavepacket is initially prepared on the left side in a single
trap. The wavepacket propagates towards the right and is split into a
coherent superposition of a state in the left arm with a state in the
right arm. Finally, applying a small offset between the currents in
the left and right wire gives rise to a phase shift between the two
wavepackets which results, after recombination, in a fringe pattern of
the longitudinal density. For a more detailed description of atom
interferometers of this type we refer to \cite{Andersson01}. Loss of coherence
between the wavepackets in the left and right arm, due to current
fluctuations in the conductors, will decrease the visibility of the
interference pattern. We will thus study the decay of the components
of the density matrix off-diagonal in the trap index. We will suppress
the trap label (L/R) in all following expressions, bearing in mind
that the left (right) index of $\rho$ will always denote a state in
the left (right) trap.

\begin{figure}[tbh]
\includegraphics[width=8cm]{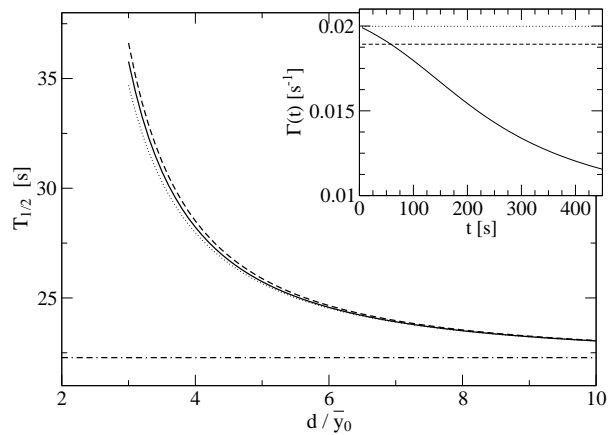}
\caption{Decay time in the double-wire trap. $T_{1/2}$ is defined 
 as the time it takes until $\rho_{00}^{00}(z,z^\prime,t)$ has decayed to
 half of its initial value at $t=0$. The graph
 shows its dependence on the wire separation $d$ 
 if all contributions in Eq.~(\ref{dwmaster}) are taken into
 account (solid line), if the cross-correlations $\beta_i$ are neglected
 (dotted line), and in the limit of $d\gg\bar{y}_0$, Eq.~(\ref{dwdecaylarged})
 (dashed line). The dash-dotted line is the $T_{1/2}$-time using the
 decay rate Eq.~(\ref{finalsingledecay}) for the single-wire
 configuration. Inset: $\Gamma(t)\equiv \ln(\rho^{00}_{00})/t$ 
 at $d=3\bar{y}_0$. The trap-chip surface distance
 taken for the plot is $\bar{y}_0 = 10\mu\textrm{m}$ and all other
 trap parameters correspond to those used in Fig.~\ref{figure3}.
 \label{figure8}}
\end{figure}

We will assume that the incoming wavepacket is initially split into a
symmetric superposition of the left and right arm. In addition we
will choose the currents in the left and right wire to be same,
$I_L=I_R$, which does not restrict the validity of the obtained result
for the decoherence effects but keeps the equations simple as the full
symmetry of the double-wire geometry is still conserved. The
dynamics of $\rho(z,z^\prime,t )$ 
is obtained from Eq.~(\ref{final2}), leading to
a matrix equation in analogy to Eq.~(\ref{master}) but with a density
matrix vector
\begin{equation}
{\bf{\rho}} = \left(\rho^{00}_{00},
\rho^{00}_{11},\rho^{11}_{00},\rho^{01}_{10},\rho^{10}_{01}
\right)\label{dwdensityvector}
\;,
\end{equation} 
where the upper index denotes the $\hat{x}$-state and the lower index
the $\hat{y}$-state of the 2d-HO. The left (right) pair of indices refers to
a state in the left (right) well.
The matrix $\tilde{A}$ is 
\begin{equation}
\tilde{A}=\left(\begin{array}{ccccc} \sum_{i=1}^4 \alpha_i & \beta_1
&\beta_2&\beta_3&\beta_4 \\ \beta_1&
\alpha_1+\alpha_3&0&\alpha_{6}&\alpha_5\\
\beta_2&0&\alpha_2+\alpha_4&\alpha_5&\alpha_{6}\\
\beta_3&\alpha_{6}&\alpha_5&\alpha_1+\alpha_4&0\\ \beta_4 &
\alpha_5&\alpha_{6}&0&\alpha_2+\alpha_3 \end{array}\right)
\label{dwmaster}.
\end{equation} 
Here, terms labelled by $\alpha_i$ are contributions of the type shown
in Fig.~\ref{figure5}a plus the corresponding term of type
Fig.~\ref{figure5}c. Terms denoted by $\beta_i$ are cross-correlations
as described by contributions shown in Fig.~\ref{figure5}b. The terms
abbreviated as $\alpha$ do not depend on the longitudinal coordinate
$z$, $z^\prime$ whereas the cross-correlations $\beta$ are functions
of $(z-z^\prime)$.

We will assume that
the extent of the wavepacket $w_z$ along the trapping well is $w_z\ll
d$. As the cross-correlations $\beta_i$ vary on a length scale of
approximately $d$, we can replace $\beta_i(z-z^\prime)$ in
Eq.~(\ref{dwmaster}) by the constant $\beta_i(0)$ \cite{overest}.

Further on we will assume that at $t=0$ the density matrix is given by: 
\begin{equation}
{\bf{\rho}}(t=0) = \left(\rho^{00}_{00}(t=0),0,0,0,0\right)\;.
\end{equation} 
Figure~\ref{figure8} shows the time $T_{1/2}$ it takes until 
$\rho^{00}_{00}(t=0)$ has decayed to half of its initial amplitude as a
function of the wire separation $d$. We assume the average currents in
the wires to be constant and we vary only the separation length $d$.
Thus the position of the trap minima given by
 \begin{eqnarray}
&y_0=\bar{y}_0&\;,\\&x_0^{L/R}=\mp\sqrt{d^2/4-\bar{y}_0^2}&\;,
\end{eqnarray}
move horizontally to the surface, leading to a wire-to-trap 
distance of
\begin{equation}
r_0=\frac{d}{\sqrt{2}}\left[1-\left(1-4\frac{y_0^2}{d^2}
\right)^{\frac{1}{2}}\right]^{\frac{1}{2}}
\;.
\end{equation}

\begin{figure}
\includegraphics[width=8.5cm]{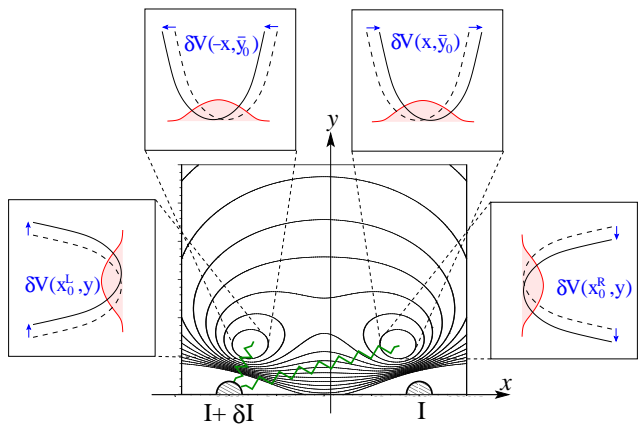}
\caption{Influence of current fluctuations $\delta I$ in the left wire
onto the potential of the trapping wells
 $ V(x,y)=\langle S|{\bm \mu}|S\rangle {\bf B}$. The potential fluctuations 
$\delta V(x,\bar{y}_0) =\delta V(-x,\bar{y}_0)$ 
are reflection symmetric about the $\hat{y}$-$\hat{z}$-plane 
for $|x-x^{L/R}_0|\ll \bar{y}_0$. Hence the fluctuations in 
$\hat{x}$-direction are
 conserving the symmetry of the trapping potential. This 
leads to positive correlations in 
$\langle \delta S_{im}(z)\delta S_{nj}(z^\prime)\rangle$ 
suppressing the decoherence. The fluctuation in 
$\hat{y}$-direction, $\delta V(x^L_0,y)$, $\delta V(x^R_0,y)$,
 show negative correlations giving rise to an increase in the
 decoherence rate.\label{figure9}}
\end{figure}

The solid line in Fig.~\ref{figure8} shows the $T_{1/2}$ time obtained
from the solution of Eq.~(\ref{master}) using Eq.~(\ref{dwmaster})
 in the approximation $\beta_i(z-z^\prime)\approx\beta_i(0)$ for
$w_z\ll d$. Cross-correlation terms $\beta_i$ play however only a
minor role for the $T_{1/2}$ time as can be seen from the dotted lines
in Fig.~\ref{figure8}, showing the case where the $\beta_i$ terms are
neglected in Eq.~(\ref{dwmaster}). The cross-correlation terms, shown
in Fig.~\ref{figure5}b, include terms describing positive correlation
between potential fluctuations,
suppressing the decoherence, but also negative correlation between potential
 fluctuations which enhance the decoherence.
Figure~\ref{figure9} shows schematically the potential fluctuations
$\delta V$ induced by a current fluctuation $\delta I$ in the left wire.

The inset of Fig.~\ref{figure8} shows the change of the decay rates over time.
Taking the full matrix Eq.~(\ref{dwmaster}) into account (solid
line) there is a change of decay rate $\Gamma_{\alpha+\beta}(t)$
 towards a slower decay
rate for $t\gg\Gamma_{\alpha+\beta}^{-1}(0)$. If the
cross-correlations are ignored, the decay rate is constant
$\Gamma_{\alpha} \approx \Gamma_{\alpha+\beta}(0)$.
 Inspection of $\tilde{A}$, given by
Eq.~(\ref{dwmaster}), shows that neglecting the terms $\beta_i$
decouples $\rho_{00}^{00}$ from the excited states leading to an
exponential decay with a single decay rate.

We finally discuss the decoherence for large distances $d$ between the
wires, i.e., $d\gg\bar{y}_0$. For $d\gg\bar{y}_0$ it 
is sufficient to take only the
influence of $\delta I_{L}$ ($\delta I_R$) onto the left (right) trap
(given by contributions of type Fig.~\ref{figure5}a) into account, as
all other contributions are strongly suppressed by orders of
$\bar{y}_0/d$ (see Table~\ref{table1}). 
Hence the matrix Eq.~(\ref{dwmaster}) decouples
$\rho^{00}_{00}$ from the excited states and the remaining equation
for $\rho^{00}_{00}$ can be solved analytically. The solution is
\begin{equation}
\rho^{00}_{00}(z,z^\prime,t) =
e^{-\Gamma t} e^{-\frac{i}{\hbar}\hat{H}
t}\rho^{00}_{00}(z,z^\prime,0) \label{dwlarged}\;,
\end{equation}
where $\hat{H}$ is the free Hamiltonian as defined in
Section \ref{sec1a}. The decay rate is given as
\begin{equation}
\Gamma=\frac{3\pi}{2\hbar^2} A_0\frac{ |x_0|^2}{d^2}
\left[1+\frac{\bar{y}_0^2}{\left(|x_0|+d/2\right)^2}\right]
\frac{1}{r_0^5}
\label{dwdecaylarged}\;.
\end{equation}
In the limit $d\gg\bar{y}_0$, Eq.~(\ref{dwdecaylarged}) reduces 
to the decay rate obtained for the single-wire configuration,
Eq.(\ref{finalsingledecay}), shown by the dash-dotted horizontal line
in Fig.~\ref{figure8}. However, there is no spatial dependence of
$\Gamma$ along the longitudinal direction $(z-z^\prime)$ in the case
of widely separated trapping wells. Plotting the
$T_{1/2}$ time for the approximation given by Eq.~(\ref{dwlarged})
(dashed line in Fig.~\ref{figure8}) using the decay rate
Eq.~(\ref{dwdecaylarged}), shows a good agreement of the approximation
with the exact solution as soon as $d$ is of the order of several
$\bar{y}_0$. Comparing all three graphs of Fig.~\ref{figure8} one can
conclude, that the increase in $T_{1/2}$ with decreasing separation
$d$ is due to geometric changes of the trap positions. An
increase of the wire-to-trap distance $r_0$ due to a decrease of $d$,
leads to an increase in the $T_{1/2}$ time as the rate of the
dominating decoherence source scales with $1/r_0^4$.

%%%%%%%%%%%%%%%%%%%%%%%%%%%%%%%%%%%%%%%%%%%%%%%%%%%%%%%%%%%%%%%%%%%%%%
\section{\label{sec3}Conclusion}
Using the density-matrix formalism, we derived an equation of motion
Eq.~(\ref{finalmess}) to describe the consequences of current fluctuations on
a cold atomic cloud in a micro-chip trap. The model allows the
description of decoherence in multiple wire traps
 \cite{Reichel02,Folman01}, as well as more complex guiding systems
required for the implementation of a beam splitter \cite{Cassettari01}
or interference experiments \cite{Hinds01,Andersson01,Haensel03}.

The atom trap was modelled as a multi-Channel 1d-waveguide, where
different channels describe different transverse modes of the
waveguide. Assuming 1d-wires on the atomic chip, we examined the
influence of current fluctuations along the wire onto the coherence of
the atom cloud in the waveguide. We found that important
contributions to the decoherence arise from transitions
to neighbouring excited states.

We used this model to examine decoherence effects for two specific
trapping configurations: the single-wire waveguide and the double-wire
waveguide. In both configurations, decoherence of the ground state was
discussed, taking processes from transitions to the first
excited states into account.

The single-wire trap showed for the ground state a decoherence rate
$\Gamma$ which scales with the wire-to-trap distance $r_0$ as
$1/r_0^4$. The potential fluctuations are correlated over a length
scale $r_0$. As a consequence the decoherence rate $\Gamma$ is a
function of the relative coordinate $z-z^\prime$. Using trap
parameters based on present experiments \cite{Folman01} we obtained
decoherence rates for $r_0=10\mu\textrm{m}$ of the order of $\Gamma
\approx 0.03\textrm{s}^{-1}$.

Extending the system to a double-wire waveguide enabled us to study
decoherence for atoms in a superposition of a state localized in the
left arm and a state localized in the right arm. This superposition is
the basic ingredient for interference experiments. Approaching the
two trapping wells, as is necessary for the splitting and merging of
the wavepackets, showed a decrease in the decoherence rate. The
decrease arises mostly due to geometrical rearrangements of the trap
minima in the system. Cross-correlation effects proved to be of minor
importance in our model. We found an explicit expression for the
decoherence rate Eq.~(\ref{dwdecaylarged}) in the limit of a
wire-to-trap distance $r_0$ much smaller than the separation of the
two wires $d$. This decoherence rate approaches the value for the
single wire waveguide, $\Gamma=A(0)$, for decreasing $r_0/d$.

The decay rates extracted in this model are small for distances
realized in present experiments \cite{Folman01}.  However we expect
that further improvement in micro-structure fabrication and trapping
techniques will decrease the trap-surface distance to scales at which
the decoherence from transitions to transverse states may have a
considerable influence on trapped atomic clouds.

\begin{acknowledgments} We would like to thank 
S.~A. Gardiner, C. Henkel, A.~E. Leanhardt, and F. Marquardt for informative
discussions, and J. Schmiedmayer for sending a copy of
Ref.~\onlinecite{Folman01} before publication. 
C.~B. would like to thank the Center for
Ultracold Atoms (MIT/Harvard) for its support and hospitality during a
one-month stay. Our work was supported by the Swiss NSF and the BBW
(COST action P5).
\end{acknowledgments}

\appendix

%%%%%%%%%%%%%%%%%%%%%%%%%%%%%%%%%%%%%%%%%%%%%%%%%%%%%%%%%%%%%%%%%%%%%%
\section{Derivation of the projected potential correlator\label{appendixA}}
This section gives a derivation of the correlation function $\langle
\delta S_{im}(z)\delta S_{nj}(z^\prime)\rangle$. A general expression
of the potential fluctuation correlator $\langle\delta
V({\bf x})\delta V({\bf x}^\prime)\rangle$ for $N$ parallel traps
will be derived, from which $\langle \delta S_{im}(z)\delta
S_{nj}(z^\prime)\rangle$ can be calculated using
Eq.~(\ref{defS}). Finally Eqs.~(\ref{SSswgeneral} -\ref{J}) and
Eq.~(\ref{SSgeneraldw}), Eq.~(\ref{defJ}) can be obtained by
specifying the geometries specializing to the single and double-wire
trap configurations.

The fluctuations of the trapping potential
$\delta V$ is induced by current noise in the wires which gives
rise to a fluctuating field $\delta {\bf B}$. The trapping
potential $\delta V$ and the magnetic trapping field $\delta{\bf B}$ 
are linked by
Eq.~(\ref{trappot}), which enables us to express 
$\langle\delta V({\bf x})\delta
V({\bf x}^\prime)\rangle$ in terms of magnetic field fluctuations:
\begin{equation}
\begin{split}
\langle \delta V({\bf x})\delta
V({\bf x}^\prime)\rangle &=
\sum_{ij}\langle
S({\bf x})|\mu_i|S({\bf x})\rangle\\
&\times\; \langle
S({\bf x}^\prime)|\mu_j|S({\bf x}^\prime)\rangle
\langle\delta B_i({\bf x})\delta
B_j({\bf x}^{\prime})\rangle\label{defVV}\;.
\end{split}
\end{equation} 
Thus the first step is to find the expression for $\langle \delta
B_i({\bf x})\delta B_j({\bf x}^\prime)\rangle$ as a
function of the current noise. We start by calculating $\delta {\bf A}$ and 
$\delta {\bf B}$ using 
\begin{eqnarray}
\delta A_i({\bf x}) &= \frac{\mu_0}{4\pi}\int
d^3\tilde{{\bf x}}\frac{\delta
j_i(\tilde{{\bf x}})}{|{\bf x}-\tilde{{\bf x}}|}
\label{defA}\;,\\
 \delta B_i({\bf x}) &= \sum_{jk}
\epsilon_{ijk}\frac{d}{dx_{j}}\delta A_k({\bf x}) \;.
\label{defB}
\end{eqnarray}
The current density in the set of 1d-wires is 
\begin{equation}
\delta {\bf j}({\bf x}) = \sum_\gamma \delta I_\gamma
(z)\delta(x-d_\gamma)\delta(y) {\bf{\hat{z}}} \label{defCDens}\;,
\end{equation}
where $d_\gamma$ denotes the $\hat{x}$-position of the $\gamma$-th
wire. The sum over $\gamma$ reflects the fact that the total current
density ${\bf j}$ 
is a sum of $M$ contributions to the current density
arising from the set of $M$ wires. We will now insert
Eq.~(\ref{defCDens}) in Eq.~(\ref{defA}) and Eq.~(\ref{defB}) to
obtain an expression for the magnetic field induced by the current in
the wire array:
\begin{equation}
\begin{split}
\delta {\bf B}({\bf x})
=&\frac{\mu_0}{4\pi}\sum_\gamma\left(
\begin{array}{c}-y\\x-d_\gamma\\0\\\end{array}\right)\\
\times & \int d\tilde{z}
\frac{\delta I_\gamma(\tilde{z})}{\left[(x-d_\gamma)^2 + y^2
+ (z -\tilde{z})^2\right]^{3/2}}\label{defB2}\;.
\end{split}
\end{equation} 
The correlation function for the current density is now assumed to be
of the form
\begin{equation}
\langle\delta I_\alpha(z)
\delta I_\beta(z^\prime)\rangle
= 4k_{\textrm{B}} T_{\textrm{eff}}\sigma A \delta(z -
z^\prime)\delta_{\alpha\beta} \label{corrCDens}
\end{equation}
for the reasons discussed before in Section \ref{sec:noise}. Using
Eq.~(\ref{corrCDens}) in combination with equations (\ref{defVV}) and
(\ref{defB2}) gives the desired relation for the
correlation function of the potential fluctuations
\begin{equation}
\begin{split}
&\langle \delta V({\bf x})\delta
V({\bf x}^\prime)\rangle = 4k_{\textrm{B}} T_{\textrm{eff}}\sigma A
\left(\frac{\mu_0}{4\pi}\right)^2\\
&\times\;\sum_\gamma\sum_{ij}\langle
S({\bf x})|\mu_i|S({\bf x})\rangle\langle
S({\bf x}^\prime)|\mu_j|S({\bf x}^\prime)\rangle\\
&\times\; Y^\gamma_{ij}({\bf x},{\bf x}^\prime)
J^{\gamma}({\bf x},{\bf x}^\prime)\;, \label{potfluct}
\end{split}
\end{equation}
where the following abbreviations have been introduced
\begin{equation}
Y^{\gamma}({\bf x},{\bf x}^\prime)=
\left(\begin{array}{ccc}y
 y^\prime &-y (x^\prime -
d_\gamma)&0\\-(x-d_\gamma)y^\prime&
(x-d_\gamma)(x^\prime
 - d_\gamma)&0\\0&0&0\\ \end{array}\right)\label{defY}\;,
\end{equation}
\begin{eqnarray}
 \nonumber
 J^{\gamma}({\bf x},{\bf x}^\prime) & = & \int\limits_{-\infty}^\infty
 d\tilde{z}\left[(x - d_\gamma)^2 + y^2 +
 (z -\tilde{z})^2\right]^{-\frac{3}{2}}\\
 \label{defJApp}
 &&\left[(x^\prime - d_{\gamma})^2 + y^{\prime 2} +
 (z^\prime -\tilde{z})^2\right]^{-\frac{3}{2}}\,.
\end{eqnarray}
Equation (\ref{defJApp}) can be further simplified if the transversal
positions $x^{(\prime)}$ and $y^{(\prime)}$ are replaced by the
position of the trap minimum $x_{0}^{\alpha}$ and $y_{0}^{\alpha}$
where $\alpha$ is the trap label denoting the trap in which the
wavefunction is localized. The replacement of the transversal
coordinates by its trap minima positions is a good approximation as
the transversal widths $w$ of the trapped atomic clouds are in general
much smaller than the wire-to-trap distance $r_0$. We have thus
reduced $J^\gamma$ to a function which now only depends on the
difference $\zeta_-=z - z^\prime$:
\begin{equation}
\begin{split}
J^{\gamma}_{\alpha\beta}(\zeta_-)=&\int\limits_{-\infty}^\infty
d\tilde{z}^\prime \left[(x_{0}^{\alpha} - d_\gamma)^2 + {y_{0}^{\alpha}}^2 +
 \left.\tilde{z}^\prime\right.^{2} \right]^{-\frac{3}{2}}\\
\times\;&\left[(x_{0}^{\beta} - d_{\gamma})^2 + {y_{0}^{\beta}}^2 +
(\zeta_- -\tilde{z}^\prime)^2\right]^{-\frac{3}{2}}
\label{defJApp2}\;,
\end{split}
\end{equation}
where we shifted the integration variable to $\tilde{z}^\prime=\tilde{z}~-~z$.
Applying the formula to the single and double wire configurations
leads to Eq.~(\ref{J}) and Eq.~(\ref{defJ}) respectively.
 
To obtain $\langle \delta V({\bf x})\delta
V({\bf x}^\prime)\rangle$ using Eq.~(\ref{potfluct}), we still need
to calculate the mean value of the atomic magnetic moment $\langle
S({\bf x})|\bm{\mu}|S({\bf x})\rangle$. We assume
that the magnetic moment follows the magnetic trapping field
adiabatically which is reasonable as long as the Larmor precession
$\omega_L=\mu_B B/\hbar$ is fast compared to the trap frequency
$\omega$. Calculating the spinor $|S({\bf x})\rangle$ for an atom
having spin $F=2$, either by considering the small corrections of the
transversal magnetic field to $B^{(z)}_{\textrm{bias}}$ perturbatively
or by calculating the rotation of the spinor as the atomic moment
follows the trapping field adiabatically, results in the following
expression for the spatial dependence of $|S({\bf x})\rangle$ for
small deviations from the trap minimum:
\begin{equation}
 |S({\bf x})\rangle = |2,2\rangle + \frac{B_x({\bf x})
 +iB_y({\bf x})}{B^{(z)}_{\textrm{bias}}} |2,1\rangle \label{F}\;.
\end{equation}
The spin states are denoted as $|F,m_F\rangle$ and the spin
quantisation axis is chosen along the $\hat{z}$-axis. Making use of
Eq.~(\ref{F}) leads to
\begin{equation}
\langle S({\bf x})|\bm{\mu}|S({\bf
x})\rangle = 2\mu_B g_F\frac{{\bf B}{({\bf
x})}}{B^{(z)}_{\textrm{bias}}} \label{mu}\;,
\end{equation}
which is an approximation to the mean value of the magnetic moment to
first order in $B_\perp/B^{(z)}_{\textrm{bias}}$. Inserting the
expression for the magnetic moment Eq.~(\ref{mu}) into Eq.~(\ref{potfluct})
 and inserting the specific magnetic trapping fields for
the single-wire trap, Eq.~(\ref{defBsw}), or the double-wire trap,
Eq.~(\ref{defBdw}), we finally obtain the results for 
$\langle \delta S_{im}(z)\delta S_{nj}(z^\prime)\rangle$ given by Eq.~(\ref{A})
for the single-wire setup and by
Eq.~(\ref{SSgeneraldw}) for the double-wire setup.


\begin{thebibliography}{00}
\bibitem{overest} Considering the opposite limit $w_z \gg d$,
we can approximate $\beta_i(z-z^\prime)\approx\beta_i
\delta(z-z^\prime)$ which results in $T_{1/2}$-times 
between the solid and dotted curves in Fig.~\ref{figure8} depending
on $z - z^\prime$.
\end{thebibliography}
\end{document}